# Tuning and Stabilizing Topological Insulator $Bi_2Se_3$ in the Intrinsic Regime by Charge Extraction with Organic Overlayers

Liang Wu[1,a)], R.M. Ireland[2,a)], M. Salehi[3], B. Cheng[1], N. Koirala[4], S. Oh[4], H. E. Katz[2,b)],

N. P. Armitage[1,c)]

1. Department of Physics & Astronomy, The Johns Hopkins University, Baltimore, MD, USA, 21218
2. Department of Material Science and Engineering, The Johns Hopkins University, Baltimore, MD, USA, 21218
3. Department of Material Science and Engineering, Rutgers, The State University of New Jersey, Piscataway, New Jersey 08854, U.S.A
4. Department of Physics & Astronomy, Rutgers, The State University of New Jersey, Piscataway, New Jersey 08854, U.S.A.

In this work, we use charge extraction via organic overlayer deposition to lower the chemical potential of topological insulator $Bi_2Se_3$ thin films into the intrinsic (bulk-insulating) regime. We demonstrate the tuning and stabilization of intrinsic topological insulators at high mobility with low-cost organic films. With the protection of the organic charge extraction layers tetrafluorotetracyanoquinodimethane(F4TCNQ) or tris(acetylacetonato)cobalt(III) ($Co(acac)_3$), the sample is stable in the atmosphere with chemical potential ~135 meV above the Dirac point (85 meV below the conduction band minimum, well within the topological insulator regime) after four months, which is an extraordinary level of environmental stability. The Co complex demonstrates the use of an organometallic for modulating TI charge density. The mobility of surface state electrons is enhanced as high as ~2000 $cm^2/Vs$. Even at room temperature, a true topologically insulating state is realized and stabilized for months' exposure to the atmosphere.

Topological insulators (TIs) are exotic bulk insulators which host helical metallic surface states on their boundaries[1,2]. Spin direction is locked to momentum and "up" and "down" spins travel in opposite directions without backscattering, which makes TIs possibly ideal platforms for future low-dissipation spintronics applications[3]. Spin plasmons generated in the topological surface states may be used for the next generation of plasmonic devices[4]. Topological surface states also display a host of interesting fundamental physics phenomena including a topological magneto-electric effect and axion electrodynamics[5, 6, 7]. Majorana fermions may be realized in the vortex core through a proximity effect between superconductors and intrinsic TIs[8]. Their non-abelian statistical nature may serve as a foundation for future fault-tolerant quantum computers[9]. All of the above phenomena can only be realized in intrinsic (i.e. bulk-insulating) TIs.

However, most TIs are either doped in the bulk or quickly lose their intrinsic properties by exposure to air[10, 11, 12]. Chemical dedoping has been shown to be effective in reducing the carrier density. Unfortunately, disorder introduced by dedoping tends to pin impurity states at the chemical potential and decrease the mobility significantly [13,14,15]. Furthermore, these chemically compensated TIs suffer from strong aging effects from the atmospheres [12]. $O_2$ was either shown to reduce the carrier density [16, 17, 18] or not to have a noticeable effect[19, 20], while $H_2O$ and CO were shown to raise the chemical potential to the conduction band and create topologically trivial Rashba states[21, 22], which interferes the transport signal from topological surface states. $H_2O$ can even serve as an electron donor for layered chalcogenide semiconductors even without prominent surface reactions [23, 24]. Therefore, an effective method to keep the sample in the bulk-insulating regime is still lacking. Gating is an effective method to lower the chemical potential in the bulk gap[26, 27, 28], however, for applications, an external high-voltage source may not be convenient and efficient in energy cost. An improved method to deplete the carrier density while keeping high mobility and preventing aging is highly necessary to facilitate applications of TIs.

Surface charge extraction (CE) is a method for modifying one semiconductor by passivating or capping it with another semiconductor[29-30]. The method relies on charge separation and energy level alignment mechanisms at interfaces. CE represents a valuable tool for the controlled and nondestructive doping of inorganic or organic semiconductors at their interfaces, especially when it cannot be easily achieved by the conventional implantation process with energetic ions. CE can effectively dedope semiconductors at the nanoscale at relatively low cost, thereby facilitating the development of hybrid electronics.

The shift of organic semiconductor (OSC) energy levels at organic/organic[29-31] and organic/metal[32-33] interfaces have been studied extensively. Organic/inorganic semiconductor junctions have been used for applications such as solar cells[34-36], field-effect transistors[30,37,38-39], diodes[40], thermoelectrics[41], and for investigating the fundamental properties of interesting electronic materials[42-46]. The tuning of the energy-level match between OSCs and metal electrodes has been achieved using both metal oxides[33, 47, 48] and organic materials as interfacial layers[37, 49]. Fermi energy pinning often results from alignment of OSC energy levels with the work functions of metals or inorganic semiconductors, specifically at their interfaces. The barrier height to charge transfer may be altered by surface modifications that shift the material work functions. However, the behavior of organic/inorganic semiconductor junctions is not perfectly predictable and requires detailed case-by-case investigation.

Regarding the topological insulator material $Bi_2Se_3$, many recent investigations have explored the shifting of energy levels by surface transfer doping/dedoping utilizing metal oxides such as ZnO[44] or $MoO_3$[45], or small-organic molecules such as F4TCNQ and TCNQ[42]. $MoO_3$ appears to a popular candidate for p-type surface charge extraction dedoping of both organic[47,48] and inorganic[43,45-46] semiconductors due to its high electron affinity and ease of processing by evaporation or from solution. Nevertheless, its relatively refractory nature makes it inconvenient for partial removal to allow for further processing such as depositing ferromagnets leads for spintronic applications[62].

In this work, we demonstrate the ability to tailor/reduce the carrier concentration of low-carrier-density $Bi_2Se_3$ films prepared by MBE[54] by utilizing various OSC layers as oxidizing surface dedopants. We show the electronic accessibility of the particular OSCs, revealing that the Fermi level of $Bi_2Se_3$ is shifted towards the Dirac point by altering $Bi_2Se_3$ carrier concentration. For these purposes we chose 2,3,5,6-tetrafluoro-7,7,8,8,-tetracyanoquinodimethane (F4TCNQ) and tris(acetylacetonato)cobalt(III) ($Co(acac)_3$) due to their commercial availability, low toxicity, and stable oxidation potential[50].

The 16QL thick (1QL ≈ 1nm) low-carrier-density $Bi_2Se_3$ films used for this study were grown on $(Bi_{1-x}In_x)_2Se_3$ buffer layers on sapphire substrates using a custom-designed molecular beam epitaxy (MBE) system at Rutgers University; details are published elsewhere[53]. DC transport, angle-resolved photoemission spectroscopy (ARPES) measurements and THz experiments have shown no signatures of bulk and trivial two-dimensional electron gas (2DEG) states [53]. $Bi_2Se_3$ samples were sealed in vacuum bags and sent to Johns Hopkins University (JHU) overnight. We use uncapped $Bi_2Se_3$ films without further cleaning. The exposure time in atmosphere is about 5 minutes after overnight in the vacuum bags and before loading into an optical cryostat for THz measurements. After finishing THz measurement on day 1, samples were kept in vacuum and were transported for charge extraction layer deposition. The transport time is about 10 min and the exposure time in atmosphere is about 2 min before loading the samples into the evaporation system. CE layers of 50 nm thickness were deposited from powders in alumina crucibles at a rate of 0.3 Å s$^{-1}$ in an Edwards thermal evaporation system at a base pressure below $3\times10^{-6}$ Torr. Based on reported temperatures over which enthalpies of sublimation of nonflurinated TCNQ and Co(acac)$_3$ were measured, we estimate that F4TCNQ and Co(acac)$_3$ were heated to between 400 K and 500 K in the deposition sources during sublimation. All deposition rates and nominal thicknesses of deposited layers were monitored by a quartz crystal microbalance. The temperature was monitored by a thermocouple placed on the backside of sapphire substrates. It was at room temperature before deposition and no heating was detected during deposition. After CE layer deposition, we immediately transported the samples back to the THz facility.

Contact-free time-domain THz spectroscopy (TDTS) in a transmission geometry was performed with a custom home-built THz spectrometer. In this technique, an approximately single-cycle picosecond pulse of light is transmitted through the sample and the substrate[54, 55]. Complex conductance is measured directly and TDTS does not require Kramers-Kronig transformation. Non-destructive TDTS is an ideal tool to study the low frequency response of these materials with both the metallic Drude response from topological surface states and an $E_{1u}$ infrared active phonon being visible [54-56].

We plot the THz conductance for 16QL bare $Bi_2Se_3$ (sample 1) at day 1, with 50 nm F4TCNQ doping layer at day 2 and at day 120 in Figure 1 (a)(b) and for bare 16QL $Bi_2Se_3$ (sample 2) at day 1, with with 50 nm Co(acac)$_3$ at day 2 and day 120 in Figure 1(c)(d). The optical conductance can be well described by a single Drude term (a Lorentzian centered at $\omega = 0$), a Drude-Lorentz term (which models the phonon) and a frequency independent real $\varepsilon_\infty$ contribution to the dielectric constant (that accounts for the effect of higher energy excitations on the low frequency physics) [54-56].

$$G(\omega) = \left[-\frac{\omega_{pD}^2}{i\omega-\Gamma_D} - \frac{i\omega\omega_{pDL}^2}{\omega_{DL}^2-\omega^2-i\omega\Gamma_{DL}} - i(\epsilon_\infty - 1)\right]\epsilon_o d. \quad (1)$$

where d is the film thickness, $\Gamma$'s are scattering rates and $\omega_p$'s are plasma frequencies. We include only include a single Drude term because no bulk and trivial 2DEG states from band bending were found in these samples and top and bottom surface states are believed to have similar carrier density and mobility[53]. From the fit, we can obtain the Drude spectral weight $\omega_{pD}^2$ and scattering rate $\Gamma_D$. Scattering rate is the half-maximum width of the Drude conductance. The Drude spectral weight (squared) is a measure of carrier density. The integrated Drude spectral weight is proportional to the area under the real part of Drude conductance $G_{D1}$. It gives the ratio of total sheet carrier density over transport effective mass.

As we can see, electron-attracting CE layers F4TCNQ and Co(acac)$_3$ both effectively lower the spectral weight of the Drude response, indicating that the surface states are depleted. The Drude conductance retains a similar scattering rate, which indicates the organic thin film layers are homogenous and induce no extra impurity scattering. This is consistent with previous spectroscopy studies, which showed that the surface states dispersion is robust to adsorbed molecules[57-60]. Nevertheless, in contrast to these works that increased the chemical potential by depositing organics[57-60], we lower it towards the Dirac point. Samples were kept in a desiccator with reduced moisture for the next four months and then re-measured by the THz spectrometer. No significant increase of spectral weight or increase of scattering were found in the THz conductance. Not only do organic CE layers decrease the carrier density, but also they stabilize the sample in the topologically intrinsic regime.

We also performed DC transport on sample 1 with F4TCNQ and sample 2 with Co(acac)$_3$ at day 30. Longitudinal resistance as a function of temperature along with standard Hall effect measurements at 5 K and at room temperature were carried out with a van der Pauw geometry in a magnetic field up to 0.6 T. The electrical contacts are made by indium wires that are manually pressed on four corners of the sample with 1cm x 1cm size. The manual pressing is sufficient enough to pierce through the capping layers assuring direct contact with the Bi$_2$Se$_3$ layer. Any residual contact resistance is eliminated by the four-point measurement. The carrier density has been extracted from the standard $R_{xy} = B/(en_{Hall})$ formula. Values for mobility were calculated based on $\mu = \sigma_{xx}/(en_{Hall})$. The control sample is a 16QL sample capped by PMMA and was measured on day 2. From Figure 2, we can see the sheet carrier density $n_{2D}$ decreases from $6.4\times10^{12}$/cm$^2$ in bare Bi$_2$Se$_3$ to $4.2\times10^{12}$/cm$^2$ in the F4TCNQ sample and to $4.0\times10^{12}$/cm$^2$ in the Co(acac)$_3$ sample while mobility is enhanced from 1320 cm$^2$/Vs to 1780 cm$^2$/Vs and 1940 cm$^2$/Vs respectively.

We fit the data to analyze the quantitative change from THz spectra. A representative fit is shown as dashed lines in Figure 1(a)(b). Fitting parameters are plotted in Figure 3 (a)(b). For one surface state, carrier density is expressed as $k_F^2/4\pi$. $n_{2D}$ is the total sheet carrier density, accounting for two nominally identical surfaces. $m^*$ is the transport effect mass defined for Dirac fermion $m^* = \hbar k_F/v_F$. For massless Dirac fermions, $n_{2D}$ and $m^*$ are related and both can be expressed as a function of $k_F$. Using the known surface state dispersion relation $E=Ak_F+Bk_F^2$ [55], the Fermi velocity can also be expressed as a function of $k_F$ by $v_F = \partial E_F/\hbar\partial k$. In the end, we have a relation between spectral weight and $k_F$ directly and we can calculate all the quantities including effective mass and chemical potential[56].

$$\frac{2}{\pi\varepsilon_0}\int G_{D1}d\omega = \omega_{pD}^2 d = \frac{n_{2D}e^2}{m^*\epsilon_0} = \frac{k_F(A+2Bk_F)e^2}{2\pi\hbar^2\epsilon_0} .(2)$$

The spectral weight is reduced, consistent with depletion of charges. The scattering rate is more or less unchanged or decreased by a small amount, likely because lowering of the chemical potential reduces the phase space for scattering. We use equation 2 to calculate $k_F$ from the spectral weight and then calculate carrier density, effective mass and chemical potential[56]. The mobility can be calculated by $\mu = e/(m^*\Gamma_D)$. The carrier density and mobility as a function of time are plotted in Fig.3(c)(d). As we can see, F4TCNQ decreased the carrier density of sample 1 from 6.7 ×$10^{12}$/$cm^2$ to 4.0 ×$10^{12}$/$cm^2$, while the mobility is enhanced from 1440 $cm^2$/Vs to 1950 $cm^2$/Vs. More importantly, F4TCNQ protects the sample from severe aging effects. For bare $Bi_2Se_3$, carrier density would increase to 10×$10^{12}$/$cm^2$, which corresponds to the conduction band minimum, within a week[12]. After four months, sample 1 with F4TCNQ has carrier density 4.5 ×$10^{12}$/$cm^2$ and mobility 1860 $cm^2$/Vs. Converting this into the chemical potential value above the Dirac point, it is still stabilized at $E_F$~140 meV (80 meV below the conduction band) after four months. F4TCNQ is known as one of the strongest hole donors/electron acceptors among organics. $Co(acac)_3$ is another very effective organic CE layer that we found to be as effective for this system as F4TCNQ. It depletes the carrier density of $Bi_2Se_3$ sample 2 from 6.9×$10^{12}$/$cm^2$ to 4.2×$10^{12}$/$cm^2$ while the mobility is enhanced from 1260 $cm^2$/Vs to 2020 $cm^2$/Vs. After four months, sample 2 with $Co(acac)_3$ has a carrier density 4.4×$10^{12}$/$cm^2$ and mobility 1980 $cm^2$/Vs. The chemical potential above the Dirac point is 135 meV (85 meV below conduction band). We also found that after a few thermal cycles and a few months, these organic film did not show visible cracks or flaking and they are still mirror-like. Therefore, we believe their adhesion to the $Bi_2Se_3$ films is excellent.

We also performed room-temperature DC transport measurements on these two samples on day 30. Even at room temperature, sample 1 with F4TCNQ has a carrier density of 4.3×$10^{12}$/$cm^2$ and mobility of 1080 $cm^2$/Vs and sample 2 with $Co(acac)_3$ has a carrier density of 4.5×$10^{12}$/$cm^2$ and mobility of 1120 $cm^2$/Vs. This shows even at room temperature the chemical potential is ~80 meV below the conduction band minimum. Thus, we realized and stabilized a room-temperature intrinsic topological insulator with high mobility. We have repeated experiments on F4TCNQ and $Co(acac)_3$ on another two samples and they show the same results. To compare with our OSC overlayers, we have also thermally evaporated $MoO_3$ on top of $Bi_2Se_3$ and performed THz spectroscopy (Figure 1(e)(f)) and DC transport (Figure 2) on these samples. We found that $MoO_3$ reduces the total sheet carrier density to 3.6 × $10^{12}$/$cm^2$, which corresponds to both surface states having chemical potential ~120 meV above the Dirac point.

Three compounds deplete the carriers similarly despite their different molecular orbital energies. The Fermi energy of isolated $Bi_2Se_3$ is reported to be about 4.9 eV below the vacuum level, as shown in Figure 4. The reason why thermally evaporated $MoO_3$ does not reduce more carrier density is likely the film quality is not as good as MBE deposition[53] and it is ex-situ. The good performance of $Co(acac)_3$ is surprising consider its high LUMO bands. While this does not indicate a thermodynamic/enthalpic driving force for the extraction of electrons from $Bi_2Se_3$ by $Co(acac)_3$, there are a number of mechanisms by which this could occur. First, the oxidizing power of solid $Co(acac)_3$ might be higher than the dissolved form. Second, the solid form might contain Co(III) species that are not fully coordinated, so individual sites of higher oxidizing power might be present. Third, the formation of an interfacial dipole between $Bi_2Se_3$ and $Co(acac)_3$ might interpose an additional driving voltage promoting electron transfer out of $Bi_2Se_3$ into $Co(acac)_3$. Finally, entropic considerations would lead to some electron transfer in spite of unfavorable enthalpy. In addition to electron transfer, the $Bi_2Se_3$-$Co(acac)_3$ interface could be rich in intermetallic chemical species that could act as electron traps, such as by the passivation of Se vacancies that are the origin of the mobile electrons.

To conclude, these organic CE layers serve as an ideal capping layer because they not only reduce the carrier density but also protect samples from degrading reaction with the atmosphere. These samples can also be further combined with conventional gating experiments to tune the chemical potential past the Dirac point dielectric layers can be directly deposited on top these organic CE layers. Such CE layers are potentially very configurable as they can be removed in small areas by acetone without decreasing the mobility in order to provide contact areas for ferromagnetic[61] or superconducting[62] leads for future spintronics and quantum computing applications or CE layers can be deposited after these leads are deposited [27]. Gapping the Dirac points by magnetic proximity effect by Co(Ni)-TCNQ with the formation of long-range magnetism would be also interesting for future spintronics devices [63]. Patterning of these bulk-insulating topological insulators may allow the realization of spin plasmonics devices [4].

**Acknowledgements.**

We thank M. Fuhrer for helpful discussions. This work was supported by the NSF DMR-1308142 with additional support by the Gordon and Betty Moore Foundation through Grant GBMF2628 to NPA at JHU and EPiQS Initiative Grant GBMF4418 to SO at Rutgers.

a) These authors contributed equally to this work.

b) hekatz@jhu.edu
c) npa@jhu.edu

Field-Effect Transistors by Overlaying Discontinuous Nano-Patches of Charge-Transfer Doping Layer on Top of Semiconducting Film.

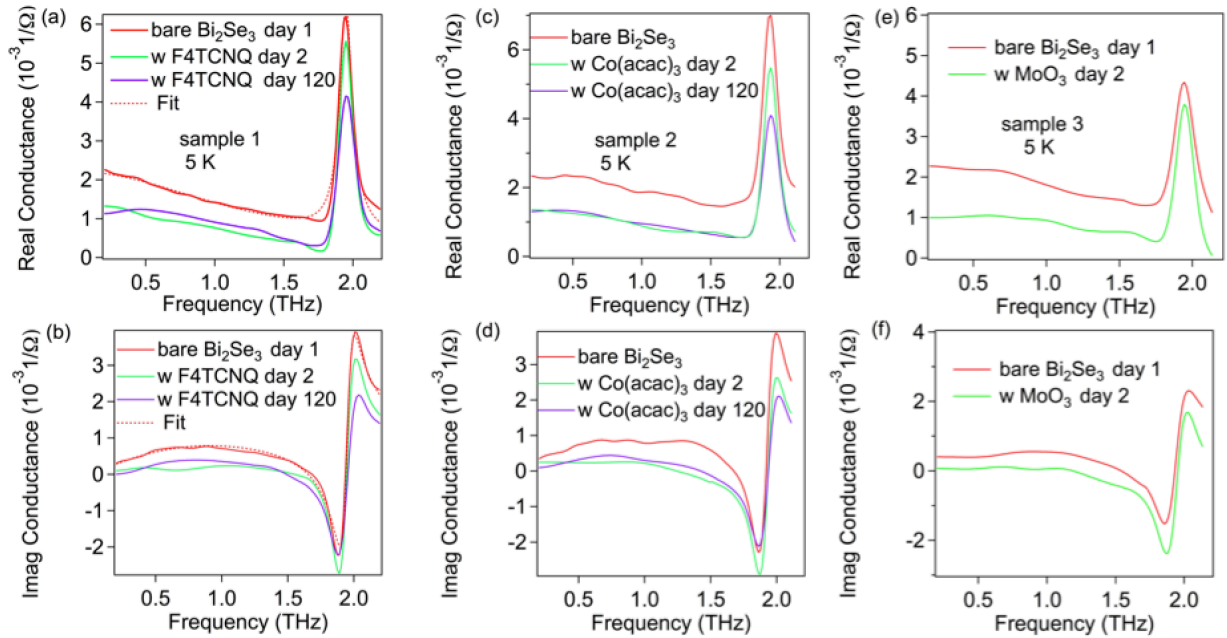

**Figure 1.** (a) Real and (b) imaginary conductance for bare 16QL $Bi_2Se_3$ (sample 1) at day 1, with 50 nm F4TCNQ doping layer at day 2 and with F4TCNQ at day 120 at 5 K. (c) Real and (d) imaginary conductance for bare 16QL $Bi_2Se_3$ (sample 2) at day 1, with 50 nm $Co(acac)_3$ doping layer at day 2 and with $Co(acac)_3$ at day 120 at 5 K. (e) Real and (f) imaginary conductance for bare 16QL $Bi_2Se_3$ (sample 3) at day 1, with 50 nm $MoO_3$ doping layer at day 2 at 5 K.

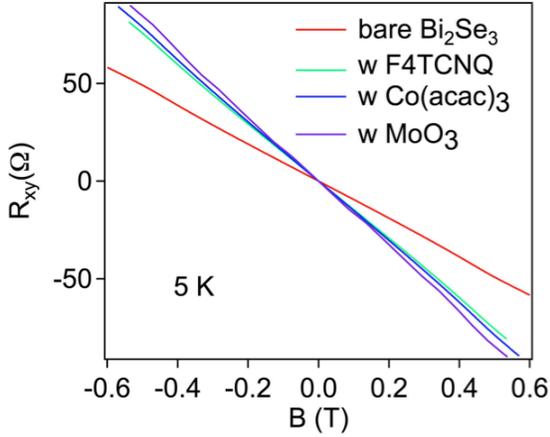

**Figure 2.** Hall resistance for 16QL bare $Bi_2Se_3$, with 50 nm F4TCNQ, with 50 nm $Co(acac)_3$ and with 50 nm $MoO_3$ at 5 K.

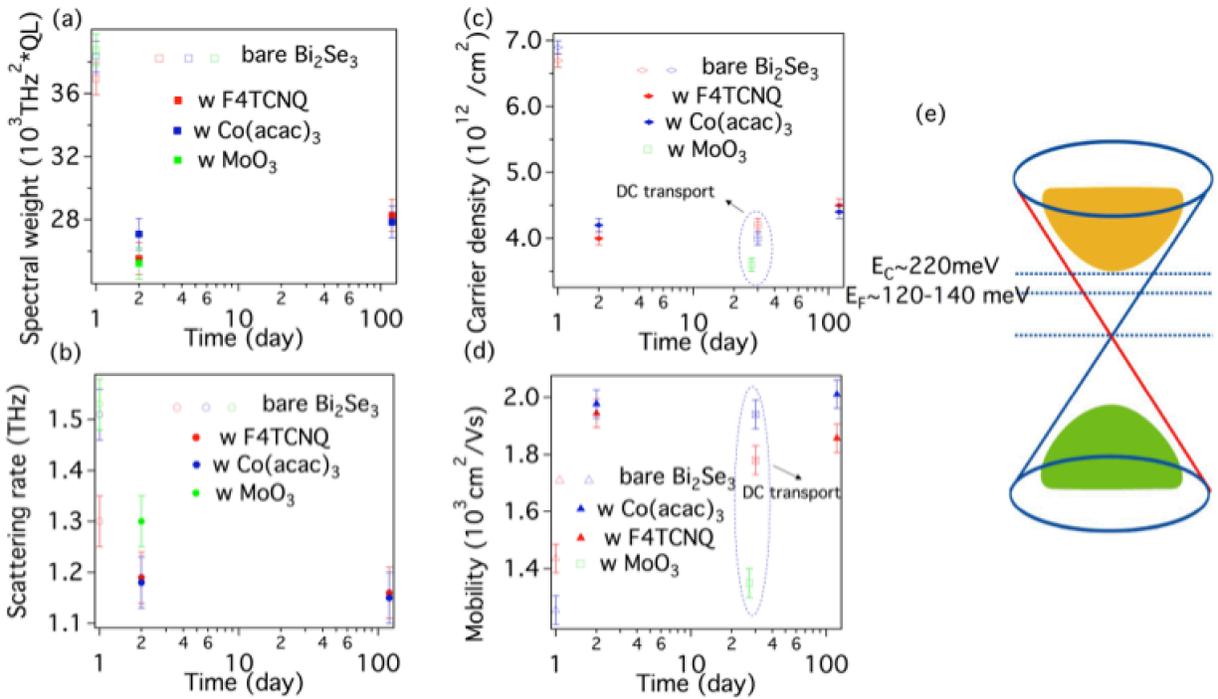

**Figure 3.** (a) Drude spectral weight (b) Drude scattering rate (c) carrier density and (d) mobility as a function of time for 16QL bare $Bi_2Se_3$, with 50 nm F4TCNQ, with 50 nm $Co(acac)_3$ and with 50 nm $MoO_3$ at 5 K. The data points marked with open squares are from DC transport. (e) Energy level diagram indicates the Fermi level of TI/OSC devices have chemical potential 80-100 meV below conduction band minimum.

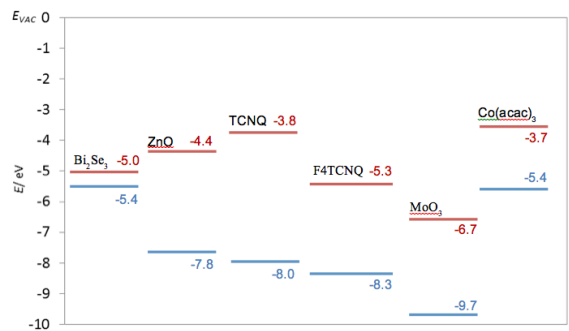

**Figure 4.** Energy level diagram with energy band edges for $Bi_2Se_3$[51], ZnO[29], 7,7,8,8-tetracyanoquinodimethane (TCNQ)[29], F4TCNQ[29], $MoO_3$[48], and redox potentials of $Co(acac)_3$[52]. Blue and red lines stand for highest occupied molecular orbital (HOMO) and lowest unoccupied molecular orbital (LUMO) bands.